\documentclass[fleqn,usenatbib]{mnras}

\usepackage{mathptmx}

\usepackage[T1]{fontenc}

\usepackage{comment}

\DeclareRobustCommand{\VAN}[3]{#2}
\let\VANthebibliography\thebibliography
\def\thebibliography{\DeclareRobustCommand{\VAN}[3]{##3}\VANthebibliography}

\usepackage{graphicx}	
\usepackage{amsmath}	
\usepackage{amssymb}	
\usepackage{ulem}

\title[Stellar Migration from Andromeda]{On Stellar Migration from Andromeda to the Milky Way}

\author[Lukas G{\"u}lzow et al.]{
Lukas G{\"u}lzow$^{1,2}$\thanks{E-mail: lukas.guelzow@kit.edu},
Malcolm Fairbairn$^{3}$,
Dominik J. Schwarz$^{2}$
\\
$^{1}$Institut für Astroteilchenphysik (IAP), Karlsruhe Institut f{\"u}r Technologie (KIT), D-76131 Karlsruhe, Germany\\
$^{2}$Fakult{\"a}t für Physik, Universit{\"a}t Bielefeld, Postfach 100131, D-33501 Bielefeld, Germany\\
$^{3}$Department of Physics, King's College London, Strand, London WC2R 2LS, UK\\
}

\date{Accepted XXX. Received YYY; in original form ZZZ}

\pubyear{2023}

\begin{document}
\label{firstpage}
\pagerange{\pageref{firstpage}--\pageref{lastpage}}
\maketitle

\begin{abstract}
Recent \textit{Gaia} observations suggest that some hypervelocity stars (HVSs) might originate from outside the Galaxy.
We ask if these HVSs could come from as far as Andromeda.
Therefore, we simulate HVSs originating in Andromeda with initial conditions based on attributes of high-velocity stars measured in the Milky Way and a simple model for the gravitational potential of Andromeda and the Milky Way.
We evaluate the validity of this scenario based on the simulation results.
While we expect the vast majority of HVSs in our Galaxy will originate here, we expect the number of stars present from Andromeda at any one time to be between twelve and $3910$, depending upon model assumptions.
Further, we analyse the properties of HVSs that are able to reach the Milky Way. We discuss whether they could be detected experimentally based on recent constraints set on the ejection rate of HVSs from the Milky Way centre.

\end{abstract}




\section{Introduction}
Hypervelocity stars (HVSs) are some of the fastest objects in the Galaxy.
Some of them exceed escape velocity and are unbound to the Milky Way gravity.
Based on recent \textit{Gaia} observations, a number of them could have extragalactic origins.
In this paper, we investigate the Andromeda galaxy as a source of these hypervelocity stars.

Hypervelocity stars are  defined as stars that have velocities of the order of $1000\, \mathrm{km\, s^{-1}}$.
They were first predicted by \citet{Hills_88}.
The first HVS in the Milky Way was discovered by \citet{Brown_2005}.
Since then, the number of known HVSs in the Milky Way keeps increasing with every new probe of the Galaxy \citep{Ginsburg_13}.
As the Milky Way escape velocity is of the same order of magnitude as the typical HVS velocity \citep{Monari_2018}, it is easy to see that they can be unbound to the Milky Way gravitational potential.
Many known HVSs move away from the Milky Way \citep{Kreuzer_2020}.
A single HVS is confidently associated with the Milky Way centre (MWC) \citep{Koposov_2020}.
The cause for such high kinetic energies is thought to be gravitational interactions between binary stars and the supermassive black hole in the MWC or other massive black holes in that region \citep{Hills_88, Brown_2010}.
In this so-called Hills mechanism, one of the stars is captured by the black hole while the other is ejected at a high velocity.
In a recent series of analyses of the \textit{Gaia} Data Release 3 (DR3) catalogue \citep{GAIA_2016, GAIA_val, GAIA_2022}, constraints have been set on the ejection rate of HVSs from the MWC caused by this mechanism \citep{Marchetti_2022, Evans_2022, Evans_20222}.
Other possible origins include the ejection of one half of a binary, caused by the supernova explosion of the other half \citep{Wang_2009}, tidal tails from dwarf galaxies passing through the Milky Way \citep{Abadi_2009, Piffl_2011} and runaway stars from the Large Magellanic Cloud \citep{Boubert_2016, Erkal_2019, Evans_2021, Lin_2023}.
\citet{Irrgang_20182} also finds that additionally another ejection mechanism is likely at play.
The known HVSs are main sequence stars with masses of the order of a few Solar masses \citep{Irrgang_2018}.

\citet{Montanari_2019} investigate multiple HVSs from the \textit{Gaia} Data Release (DR2) catalogue \citep{GAIA_2016, GAIA_2018}.
They filter the data according to Galactocentric velocity as well as the probability of being unbound from the Milky Way.
The probabilities for the HVSs to be unbound are provided by \citet{Marchetti_2018}.
They find 20 HVSs with a probability $>80$ per cent of being unbound.
13 out of these 20 HVSs have trajectories that point towards the Milky Way disc.
This indicates that their origin is not within the Milky Way, but outside of it.
The trajectories of HVSs that originate inside the Milky Way characteristically point away from it.
\citet{Montanari_2019} argue that they may originate in globular clusters and dwarf galaxies surrounding the Milky Way, or even in the Andromeda Galaxy.
While \citet{Montanari_2019} focus on dwarf galaxies, this paper will focus on Andromeda as a possible origin. Previous work on this topic has been done by \citet{Sherwin_2008}.

We expect Andromeda to eject HVSs at a much higher rate than its satellites.
Compared to the dwarf galaxies in the immediate neighbourhood of the Milky Way, Andromeda and its satellites are much further away.
For this reason, we neglect  Andromeda's satellites and only consider HVSs ejected from the inner region of Andromeda.

We simulate a dynamical model of the gravitational system of Andromeda and the Milky Way in \textsc{Python} \citep{astropy:2013,astropy:2018, 2020SciPy-NMeth}.
We randomly generate initial positions near the centre of Andromeda from which the HVS trajectories start.
For the initial velocity vectors, we first take isotropically distributed directions.
Then, we randomly generate velocity magnitudes within boundaries based on Andromeda's escape velocity \citep{Kafle_2018} and the distribution of star velocities in the Milky Way \citep{Marchetti_2021}.
With this information, we calculate the trajectories of the ejected HVSs.

We found that it is possible for them to reach the Milky Way.
We approximated the amount of Andromeda HVSs in the Milky Way at present time and analysed their position and velocity properties.
In addition, we discuss whether it is possible to detect them.

In Section~\ref{sec:Methods}, we explain the utilised models for the simulation of the Milky Way, Andromeda and the HVS trajectories in the system as well as the generation of initial conditions.
Section~\ref{sec:Results} discusses the results of the simulation, the properties of the HVSs close to the Milky Way and compares them to data from the \textit{Gaia} DR3 catalogue.
Finally, we draw our conclusions in Section~\ref{sec:Concl}.

\section{Model and Simulations}
\label{sec:Methods} 

There are three main components to the simulation of HVS trajectories in the system of the Milky Way and Andromeda: First, we calculate the relative motion of the two galaxies in Section \ref{sec:Andr_traj}. Secondly, we model the gravitational potential of the entire system in Section \ref{sec:Potential}. And finally, we generate initial conditions and use the equations of motions to simulate HVS trajectories in Section \ref{sec:maths}.

\subsection{Andromeda trajectory}
\label{sec:Andr_traj}
The system of the Milky Way and Andromeda is a dynamical one.
Before we can determine any HVS trajectories, we calculate the trajectory of Andromeda in order to have an accurate dynamical model. 
Since the two galaxies do not collide during the relevant time frame for the propagation of HVSs, we describe both galaxies by point masses for the purpose of calculating Andromeda's trajectory.
Other gravitational influences in- and outside the Local Group are neglected (including cosmological expansion).
In addition, we assume a present age of the Universe of $t_0 = 13.8$ Gyr.

We choose the coordinate system for this calculation as well as in the HVS trajectory simulation so that the MWC lies in the origin and is at rest.
Andromeda lies on the $x$-axis and at $y=z=0$ at present time.
Additionally, the $y$-axis is chosen so that the Sun lies within the $xy$-plane to make the transformation to Galactic coordinates more convenient.
The definition of the axis orientations only holds at present time, however,  the coordinate system does not change in time. 

\begin{figure}
	\includegraphics[width=\columnwidth]{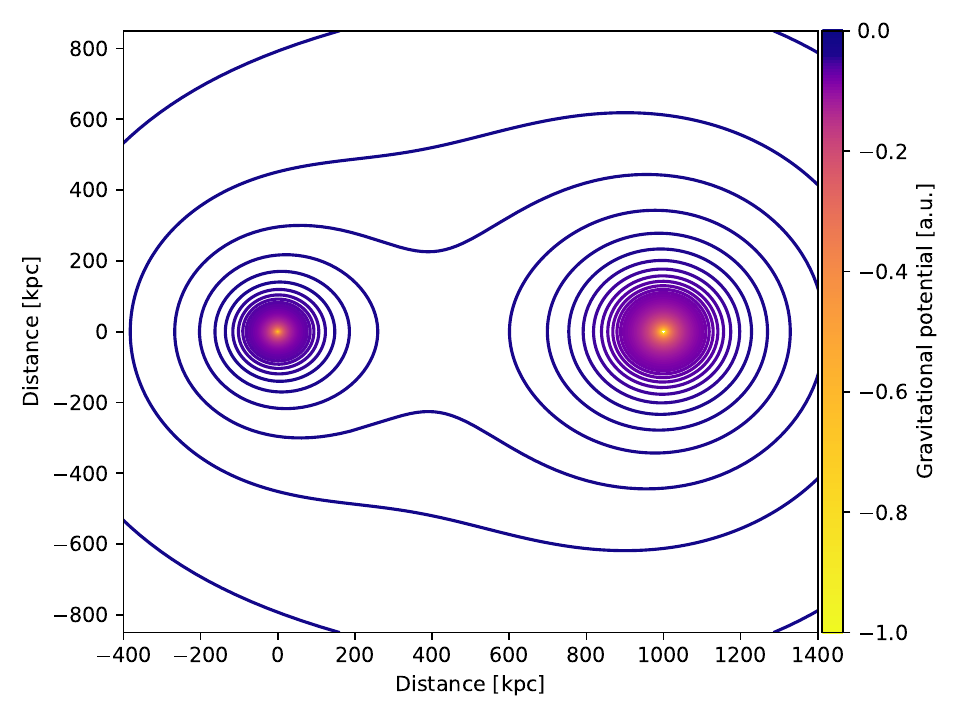}
    \caption{Contour plot of the combined potential of the Milky Way (at the origin of the coordinate system) and Andromeda galaxies in the half-mass scenario in arbitrary units at $t=10\,\mathrm{Gyr}$ after the Big Bang. The potential is computed according to our model for galaxy mass distribution, with the Plummer model for baryonic mass and the NFW model for the dark matter haloes. The potential has a saddle point between the two
    galaxies.}
    \label{fig:potential}
\end{figure}

The acceleration vector $\mathbfit{a}_\text{And}$ acting on Andromeda follows from the two-body problem,
\begin{equation}
\begin{aligned}
\mathbfit{a}_\text{And}(\mathbfit{r})=-G\frac{M_\text{MW} + M_\text{And}}{|\mathbfit{r}|^3}\,\mathbfit{r},
\end{aligned}
\label{a_and}
\end{equation}
with $M_\text{MW}$ and $M_\text{And}$ as the respective total masses of each galaxy and $\mathbfit{r}$ the position vector of Andromeda with respect to the MWC.
Given the current position and velocity coordinates of Andromeda in the Galactocentric rest frame from \citet{van_der_Marel_2012},
\begin{equation}
\begin{aligned}
    \mathbfit{r}_\text{M31} &= \begin{pmatrix}
           -378.9 \\
           612.7 \\
           -283.1 \\
         \end{pmatrix}\,\mathrm{kpc}
    \qquad
    \text{and}\\
    \mathbfit{v}_\text{M31} &= \begin{pmatrix}
           66.1\pm 26.7 \\
           -76.3\pm 19.0 \\
           45.1\pm 26.5 \\
         \end{pmatrix}\,\mathrm{km\, s^{-1}},
\end{aligned}
\end{equation}
we can determine Andromeda's trajectory at any point in time with Equation~(\ref{a_and}) and the equations of motion established in Section~\ref{sec:maths}.

\subsection{Gravitational potential}
\label{sec:Potential}
The total acceleration acting on an object at any position in the system at any time is given by the gravitational potential.
To find the total potential, we model the mass distribution of the two galaxies.
We combine two different density profiles for the mass term of a single galaxy.
The Navarro--Frenk--White (NFW) profile \citep{Navarro_1996} is used for the dark matter halo of a galaxy, making up the bulk of its mass.
It is a spherically symmetric halo model based on N-body simulations in a cold dark matter scenario.
It follows the mass density relation
\begin{align}
\rho_\text{NFW}(r)&=\frac{\rho}{\frac{r}{R_\text{s}} \left(1 + \frac{r}{R_\text{s}}\right)^2}.
\end{align}
Here, $\rho$ is the characteristic density and $R_\text{s}$ denotes the scale radius which both differ for every halo.

We add a second component of mass resulting from the Plummer density profile \citep{Plummer_1911}.
The Plummer profile approximately describes the distribution of baryonic matter in a galaxy.
It is a spherically symmetric model as well which means it is at its most accurate near the centre of a galaxy.
In the Plummer model, the mass density reads
\begin{align}
\rho_\text{P}(r)&=\frac{3M_{\text{B}}}{4\pi\cdot R_\text{P}^3}\left(1 + \frac{r^2}{R_\text{P}^2}\right)^{-\frac{5}{2}}.
\label{Plummer}
\end{align}
Here, $M_{\text{B}}$ is the total baryonic mass of the galaxy and $R_\text{P}$ denotes the Plummer radius which determines the core radius of the galaxy.

The total acceleration $\mathbfit{a}_\text{tot}$ on a HVS caused by the gravitational influence of both galaxies follows by inserting these two mass components into Newton's law of gravitation.
$\mathbfit{a}_\text{tot}$ is defined by the three spacial components of $\mathbfit{a}_\text{tot}=(a_\text{tot,x}, a_\text{tot,y}, a_\text{tot,z})$ with
\begin{equation}
\begin{aligned}
a_\text{tot,i}= - 4\pi G\biggl[&\rho_\text{1} R_\text{s,1}^3\left(\mathrm{ln}\left(\frac{R_\text{s,1}+|\mathbfit{r}_1\,|}{R_\text{s,1}}\right) - \frac{|\mathbfit{r}_1\,|}{R_\text{s,1}+|\mathbfit{r}_1\,|}\right)\frac{r_\text{1,i}}{|\mathbfit{r}_1\,|^3}\\
+ \,&\rho_2 R_\text{s,2}^3\left(\mathrm{ln}\left(\frac{R_\text{s,2}+|\mathbfit{r}_2|}{R_\text{s,2}}\right) - \frac{|\mathbfit{r}_2|}{R_\text{s,2}+|\mathbfit{r}_2|}\right)\frac{r_{2,i}}{|\mathbfit{r}_2|^3}\\
+ \,&\frac{M_\text{B,1}}{4\pi} \frac{r_\text{1,i}}{\left(|\mathbfit{r}_1\,|^2+R_\text{P}^2\right)^{3/2}}\\
+ \,&\frac{M_\text{B,2}}{4\pi} \frac{r_{2,i}}{\left(|\mathbfit{r}_2|^2+R_\text{P}^2\right)^{3/2}}\biggr].
\label{eq:a_tot}
\end{aligned}
\end{equation}
Here, the top two lines correspond to the NFW profile and the bottom two lines correspond to the Plummer profile of the two galaxies.
$r_\text{1,i}$ is the $i$-component of the position vector of the HVS with respect to the MWC $\mathbfit{r}_1$, and $r_{2,i}$ is the $i$-component of the position vector of the HVS with respect to the centre of Andromeda $\mathbfit{r}_2$.
Analogously, the indices 1 and 2 respectively refer to the Milky Way and Andromeda on characteristic density $\rho$, scale radius $R_\text{s}$ and $M_\text{B}$. 
$\rho$ and $M_{\text{B}}$ both depend on the total mass of the corresponding galaxy.
$R_s$ is set to $20\, \mathrm{kpc}$ for both galaxies since their virial radii are both of the order of $R_\text{vir}=200\, \mathrm{kpc}$ \citep{Tamm_2012, Nuza_2014}.
The scale and virial radii are related by
\begin{align}
R_s=\frac{R_{\text{vir}}}{c}.
\end{align}
$c$ is a concentration parameter which has a typical value of $c=10$ for galaxies like the Milky Way and Andromeda.
We set $R_\text{P}=4\,\mathrm{kpc}$ for both galaxies because it results in a distribution of baryonic matter that is mainly concentrated within $4\,\mathrm{kpc}$ of their centres.

Due to varying values in the estimation of the total masses of both the Milky Way and Andromeda, we investigate two distinct scenarios in this paper:
\begin{enumerate}
\item $M_\text{MW}=M_\text{And}=0.8\times 10^{12}\,\mathrm{M}_{\sun}$ (Equal-mass scenario) \label{i}
\item $M_\text{MW}=0.615\times 10^{12}\,\mathrm{M}_{\sun} \approx 0.54\, M_\text{And}$ (Half-mass scenario).\label{ii}
\end{enumerate}
In scenario~\ref{i}, we assume both galaxies to each have an approximate total mass of $0.8\times 10^{12}\,\mathrm{M}_{\sun}$.
This is supported by \citet{Kafle_2014}, \citet{Kafle_2018}, \citet{Karukes_2020}, \citet{Cautun_2020} and \citet{Correa_Magnus_2021}.
In scenario~\ref{ii}, we assume a mass ratio $M_\text{MW}/M_\text{And}\approx 0.54$ between the two galaxies.
This scenario is supported by \citet{Watkins_2010}, \citet{Pe_arrubia_2014} and \citet{Marchetti_2021}.
A 2D representation of the gravitational potential of the two galaxies in the half-mass scenario is shown in \autoref{fig:potential}.

\subsection{Modelling of hypervelocity star trajectories}
\label{sec:maths}

To calculate a single HVS trajectory, we need to solve the equations of motion.
In this case, the equations of motion are ordinary differential equations (ODEs) for the position $\mathbfit{r}$ and velocity $\mathbfit{v}$ of the star
\begin{equation}
\begin{aligned}
\frac{d\mathbfit{r}\,(t)}{dt}&=\dot{\mathbfit{r}}\,(t)=\mathbfit{v}\,(t) \\
\frac{d\mathbfit{v}\,(t)}{dt}&= \dot{\mathbfit{v}}(t,\mathbfit{r}\,(t))=\mathbfit{a}\,(t,\mathbfit{r}\,(t)).
\label{eq:DEs}
\end{aligned}
\end{equation}
In three dimensions, we have a total of six equations, one for each of the three components of the position vector $\mathbfit{r}(t)$ and velocity vector $\mathbfit{v}(t)$.
For each of the equations, we need a set of initial conditions with three components for both position and velocity.
They are generated randomly for every HVS with weighting conditions determining the range in which they are likely to be found.
In addition, we randomly generate an initial send-off time within the time interval $\left[10,13\right]\,\mathrm{Gyr}$ after the Big Bang for each HVS.
Outside of this interval, there is only a negligible amount of HVSs that is able to reach the Milky Way at present time.

The initial radial distance from the centre of Andromeda $r$ is determined by rearranging the Plummer model density relation in Equation\ (\ref{Plummer}).
We find the radius-density relation
\begin{align}
r(\rho_\mathrm{P,2})= R_\mathrm{P,2} \sqrt{\left(\frac{3M_\text{B,2}}{4\pi R_\mathrm{P,2}^3 \rho_\mathrm{P,2}}\right)^{\frac{2}{5}}-1}.
\label{P_density}
\end{align}
Now, we randomly generate a density value between the Plummer model maximum density \mbox{$\rho_\mathrm{P,2}(r=0)$} and $\rho_\mathrm{P,2}=0$ and substitute it in Equation\ (\ref{P_density}) to find the corresponding radius.
This results in more initial radial coordinates towards the centre of Andromeda where more stars are located.
The azimuthal angle $\varphi$ and the inclination angle $\theta$ are randomly generated to give positions isotropically distributed around the centre of Andromeda.

In order to generate realistic initial velocity vectors for HVS, we employ a mathematical theorem from extreme value theory, called the Pickands-Balkema-de Haan theorem \citep{Balkema1974, Pickands1975}. It deals with the statistical properties of extreme quantities. The theorem gives us access to the upper tail of an unknown distribution of a random variable if we assume that the distribution is continuous (which should be the case for the velocity distribution of stars).
The theorem states that the tail above a fixed threshold is then well approximated by either an exponential distribution or a Pareto distribution.
While the exponential distribution is specified by a scale and location parameter only, the Pareto distribution has an additional shape parameter.
We show below that HVS in the Milky Way are indeed described very well by an exponential distribution (we also tested the Pareto distribution which improves the fit only marginally and thus we decided to go with the simpler set of assumptions). 
Based on this theorem, we characterise the velocity distribution of HVSs that exceed the escape velocity of Andromeda with an exponential distribution.
At the same time, the extreme values are not sensitive to the exact form of the total distribution.

\begin{figure}
	\includegraphics[width=\columnwidth]{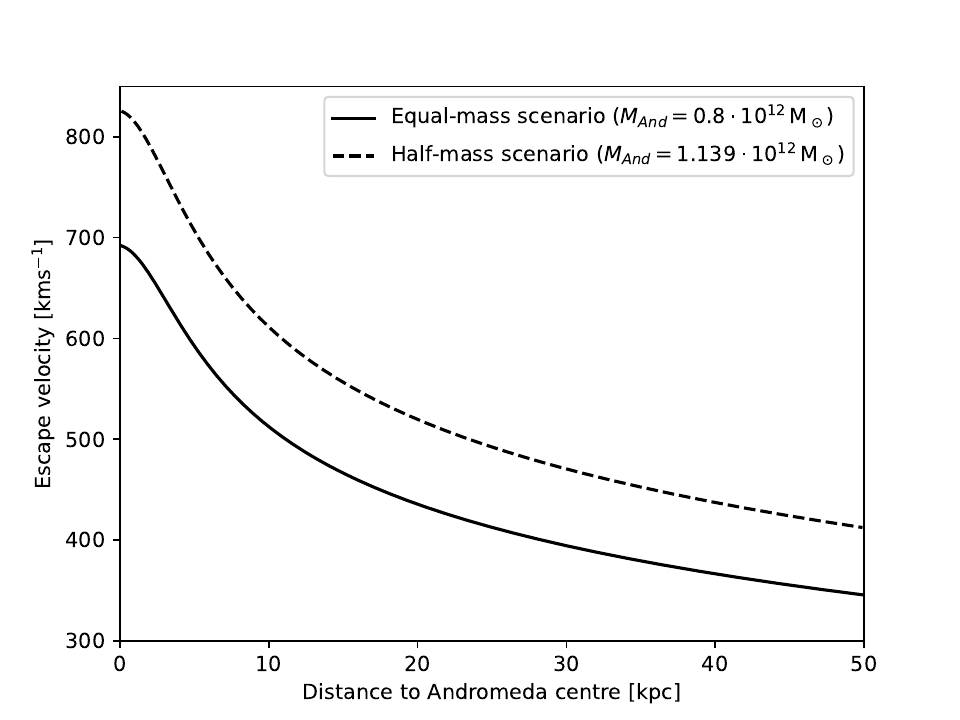}
    \caption{Escape velocity of the Andromeda Galaxy versus distance to its centre for both mass scenarios. The graph is based on the gravitational potential arising from the NFW and Plummer models used in the simulation calculations in Section~\ref{sec:Potential}.}
    \label{fig:Escape-velocity}
\end{figure}

\begin{figure}
    \includegraphics[width=\columnwidth]{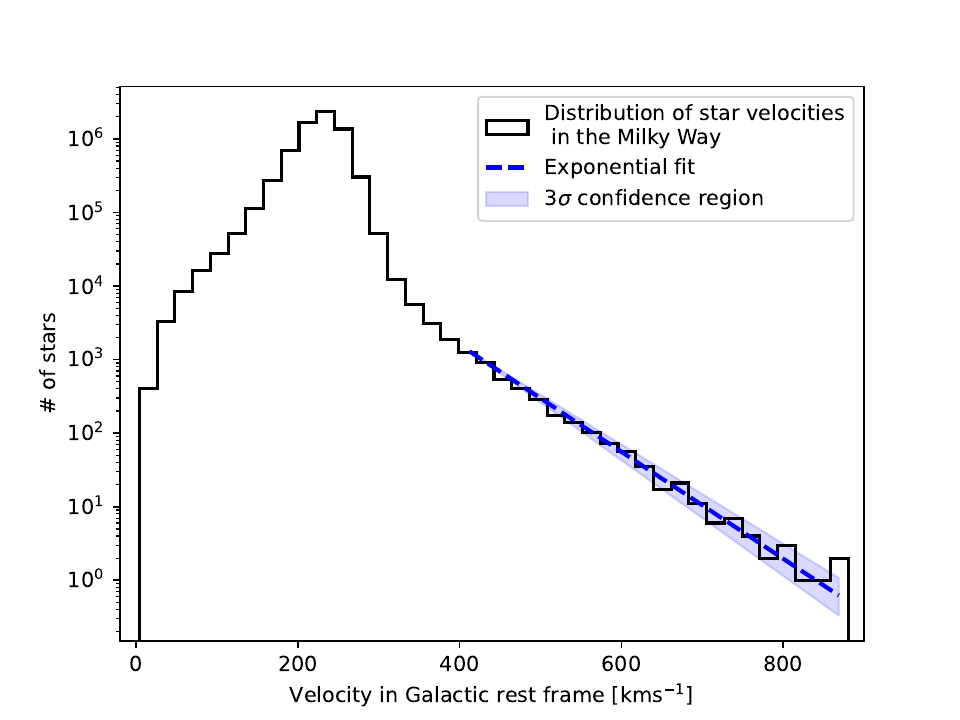}	
    \caption{Velocity distribution of all Milky Way stars in \textit{Gaia} eDR3 with sufficient information to perform velocity analysis \citep{Marchetti_2021}. Fitted with an exponential function $f(v)$ in blue in the interval $v=[400,900]\,\mathrm{km\, s^{-1}}$ in order to generate realistic initial conditions for HVSs near the centre of Andromeda. The $3\sigma$ confidence region is shown for the fit in light blue and the best-fitting values are displayed in \autoref{tab:Marchetti-fit}.}
    \label{fig:Marchetti}
\end{figure}

To find a reasonable exponential distribution, we use data from the analysis of star velocities in the Milky Way provided by \citet{Marchetti_2021}.
For the threshold of the distribution, we calculate the escape velocity of Andromeda depending on the distance to its centre according to our model.
This is shown in \autoref{fig:Escape-velocity}.
We reproduce the plot from \citet{Marchetti_2021} showing the distribution of velocities of eligible stars in the \textit{Gaia} eDR3 data catalogue \citep{GAIA_2016, GAIA_2020, GAIA_val} in \autoref{fig:Marchetti}.
We fit an exponential function
\begin{align}
f(v)=\frac{A}{\sigma_v} \exp \left( -\frac{v-v_0}{\sigma_v} \right)
\label{eq:fit-function}
\end{align}
to the high velocity tail with the fit parameters velocity scale $\sigma_v$ and amplitude $A$.
The fit is shown in blue in \autoref{fig:Marchetti}.
The fit result and the best-fitting parameters are displayed in \autoref{tab:Marchetti-fit}.
The function $P(v)=A^{-1}f(v)$ serves as the probability density function for a Milky Way star to have velocity $v$ under the condition that $v>v_0= 400\,\mathrm{km\, s^{-1}}$.
The cut-off $v_0$ is also called the location parameter of the distribution.
Picking a different cut-off in the tail of the distribution does not alter the shape parameter $\sigma_v$, but merely changes the overall normalisation $A$.

We now assume that we can adopt this present day velocity distribution of the Milky Way to Andromeda at earlier times.
We believe that, at least for the equal-mass scenario, this is a sensible assumption.

\begin{table}
	\centering
	\caption{Best-fitting parameters for the Milky Way velocity distribution in \autoref{fig:Marchetti}.}
	\label{tab:Marchetti-fit}
	\begin{tabular}{cccc} 
		\hline
		Velocity scale $\sigma_v$ $\left[\mathrm{km\, s^{-1}}\right]$ & Amplitude $A$ $\left[\mathrm{km\, s^{-1}}\right]$ & $\chi^2$ & d.o.f.\\
		\hline
		\hline
		$59.79\pm 1.27$ & $94778.15\pm 1924.41$ & 29.03 & 18\\
	\end{tabular}
\end{table}

We take the cumulative distribution function $F(v)$ corresponding to $P(v)$
\begin{align}
F(v)= 1 - \exp\left(-\frac{v-v_0}{\sigma_v} \right) \quad \text{for} \quad v\geq v_0.
\label{eq:CDF}
\end{align}
Then, we rearrange it for $v$ and substitute the escape velocity $v_\text{esc}(r)$ of Andromeda at the given position $r$ for $v_0$
\begin{align}
v\left[v_\text{esc}(r), F(v)\right]=v_\text{esc}(r)- \sigma_v \left[\mathrm{ln}(1 - F(v))\right].
\label{eq:inivel}
\end{align}
By generating a random number between 0 and 1 and substituting it for $F(v)$ in equation (\ref{eq:inivel}), we find the initial velocity magnitude $v$.

We note that the \textit{Gaia} eDR3 high-velocity star selection from \citet{Marchetti_2021} may be biased and affected by spurious measurements. For a detailed discussion of the catalogue see \citet{Marchetti_2021}.
We also do not make the assumption that the velocities of HVSs, ejected from near the centre of Andromeda, intrinsically follow the distribution of high-velocity stars in \autoref{fig:Marchetti}.
We only use the exponential fit to the distribution as a reasonable function to model the upper tail of the unknown, assumedly continuous, distribution of HVS velocities according to the Pickands-Balkeema-de Haan theorem.
\citet{Sherwin_2008} use a power law distribution for the HVS velocities, based on the ejection mechanism.
This is a special case of the Pareto distribution and, as such, also falls under the Pickands-Balkema-de Haan theorem.
Plugging their velocity distribution into our model does not change the results significantly which supports the validity of our method (see Section \ref{sec:Results}).

The angular components of the initial velocity vector are generated so that the directions are isotropically distributed.
Since we generate the initial conditions with respect to the centre of Andromeda, we add Andromeda's position and velocity at the time of ejection to the initial vectors to find the initial conditions with respect to the MWC.

We can now use the equations of motion (equation~(\ref{eq:DEs})) to calculate the trajectory of a HVS by plugging in its initial position and velocity as well as the acceleration from equation~(\ref{eq:a_tot}).

We use \textsc{Python} to model the gravitational system of the Milky Way and Andromeda galaxies.
Specifically, we utilise the \textsc{scipy} \citep{2020SciPy-NMeth}, \textsc{numpy} \citep{Harris_2020} and \textsc{astropy} \citep{astropy:2013,astropy:2018} libraries.
The equations of motion are solved in the simulation code by the \textsc{ODEint} function from the \textsc{scipy} library.
It integrates provided acceleration and velocity components to find position and velocity components, respectively. 
\textsc{ODEint} utilises the Runge-Kutta method \citep{Butcher_RK, Hairer_RK}.
\section{Result Discussion}
\label{sec:Results}
We structure the discussion in the following way: First, we analyse the properties of the simulated HVS in Section \ref{sec:HVS-properties}. Then, we compare these properties with \textit{Gaia} observations in Section \ref{sec:Gaia}. At the end, we estimate the amount of Andromeda HVSs in the Milky Way at present time and discuss if it is possible to observe them in Section \ref{sec:HVS-amount}.

\subsection{Properties of hypervelocity stars from Andromeda}
\label{sec:HVS-properties}
For every HVS, the simulation produces data at the end of the calculated trajectory at present time $t_0=13.8\,\mathrm{Gyr}$.
We also store kinematic data at its minimum distance to the MWC.
This data set includes the initial send-off time as well as the two positions in Galactic coordinates.
Due to the isotropic distribution of initial velocity directions, the majority of simulated HVSs naturally move away from the Galaxy.
We need to calculate a large number of trajectories to find a significant amount of HVSs with trajectories pointing towards the Milky Way. 

It is important to note that the amount of simulated HVS is not to be interpreted as representative of the true amount of HVSs originating in Andromeda.
Rather, it is a sufficiently large amount of data to analyse.
The various assumptions and approximations made in our model are certainly the most important source of uncertainty for the results.
Thus, we refrain from a detailed analysis of numerical errors, which are certainly much smaller, as we are relying on well established algorithms.

For each mass scenario, we calculate a total of  $1.8 \times 10^7$ trajectories with randomly generated send-off times.
We apply a filter to only consider HVSs in our analysis that reach a radius 
\begin{equation}
r\leq r_\text{Filter}=50\,\mathrm{kpc}
\end{equation}
around the MWC.
We find that about $0.08$ per cent of simulated trajectories fulfil this criterion at any point during their travel time.
The amount of results for simulated HVSs within $r_\text{Filter}$ of the MWC at present time is about one order of magnitude lower at, respectively, $0.013$ and $0.011$ per cent for the equal- and half-mass scenarios.
The exact amount of results for each mass scenario is presented in \autoref{tab:resultdist}.
The reason for the significant difference in the amount of present time results between the two scenarios is Andromeda's mass.
In the half-mass scenario, HVSs require a higher escape velocity due to Andromeda's larger mass.
This causes some the slowest HVSs that are sent off at the earliest possible times to still be too fast to remain in the Milky Way at present time.
In addition, the Milky Way's larger mass in the equal-mass scenario attracts the HVSs more strongly than in the half-mass scenario.

\begin{table}
	\centering
	\caption{Number and fraction of the total amount of simulated HVSs that reach a radius of $r_\text{Filter}$ around the MWC for the two mass scenarios and result categories.}
	\label{tab:resultdist}
	\begin{tabular}{cccc} 
		\hline
		Mass scenario & Category & HVSs within $r_\text{Filter}$ & Fraction\\
		\hline
		\hline
		Equal & Min. distance & 13998 & 0.00078\\
		Equal & Present time & 2346 & 0.00013\\
		\hline
		Half & Min. distance & 13788 & 0.00077\\
		Half & Present time & 1891 & 0.00011\
	\end{tabular}
\end{table}

\begin{figure}
	\includegraphics[width=\columnwidth]{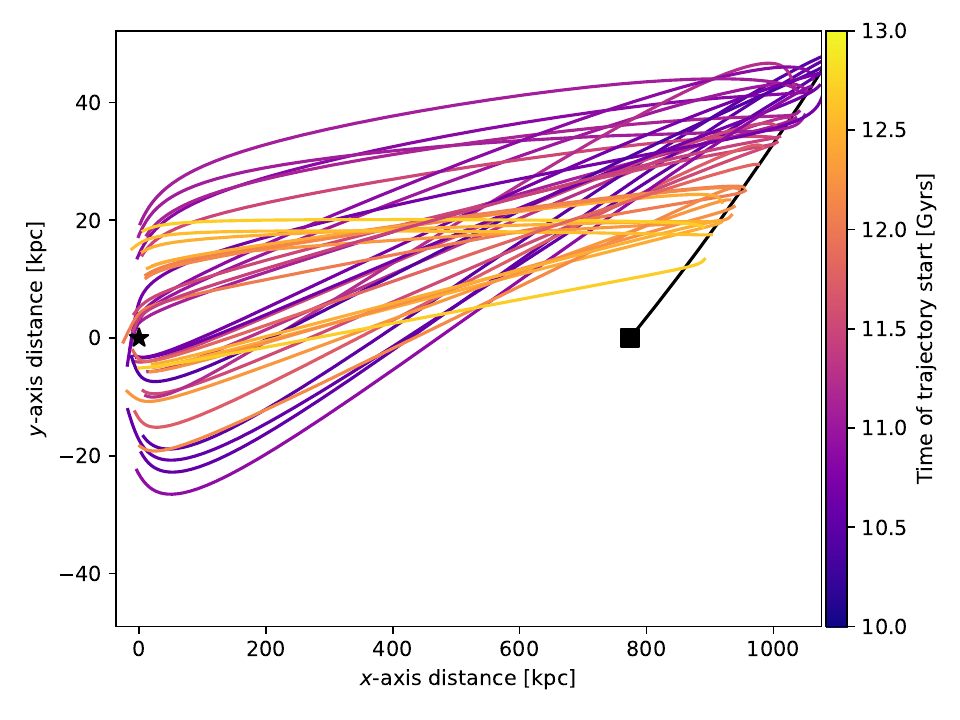}
	\includegraphics[width=\columnwidth]{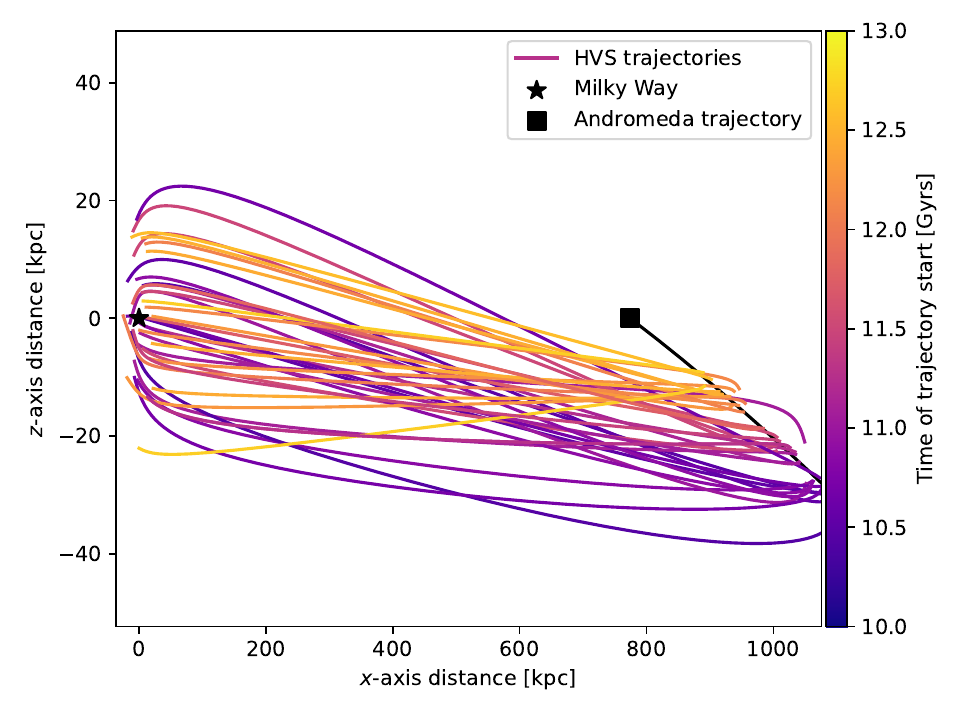}
    \caption{Projection plots of 38 full HVS trajectories as calculated by the simulation in the equal-mass scenario for the purpose of visualisation. Chosen by hand with the condition of being within $25\,\mathrm{kpc}$ of the Milky Way centre at present time $t_0=13.8\,\mathrm{Gyrs}$. Colourmapped according to the time of the start of the trajectory. Coordinate system is chosen so that, at present time, Andromeda lies on the $x$-axis and at $y=z=0$ and that the MWC and the Sun lie in the $xy$-plane. \textbf{Top:} Projection on to $xy$-plane. \textbf{Bottom:} Projection on to $xz$ -plane.}
    \label{fig:visual}
\end{figure}

\autoref{fig:visual} visualises a small, handpicked number of HVS trajectories that come especially close to the MWC in projection graphs in the $xy$- and $xz$-planes of the coordinate system used in the simulation (see Section~\ref{sec:Andr_traj}).
The $y$-axis is strongly compressed in order to show the differences between the trajectories in more detail.
The different HVS trajectory start times are visible as a colour gradient along the trajectory of Andromeda.
All shown HVS trajectories  as well as Andromeda's end at present time.

\begin{figure}
	\includegraphics[width=\columnwidth]{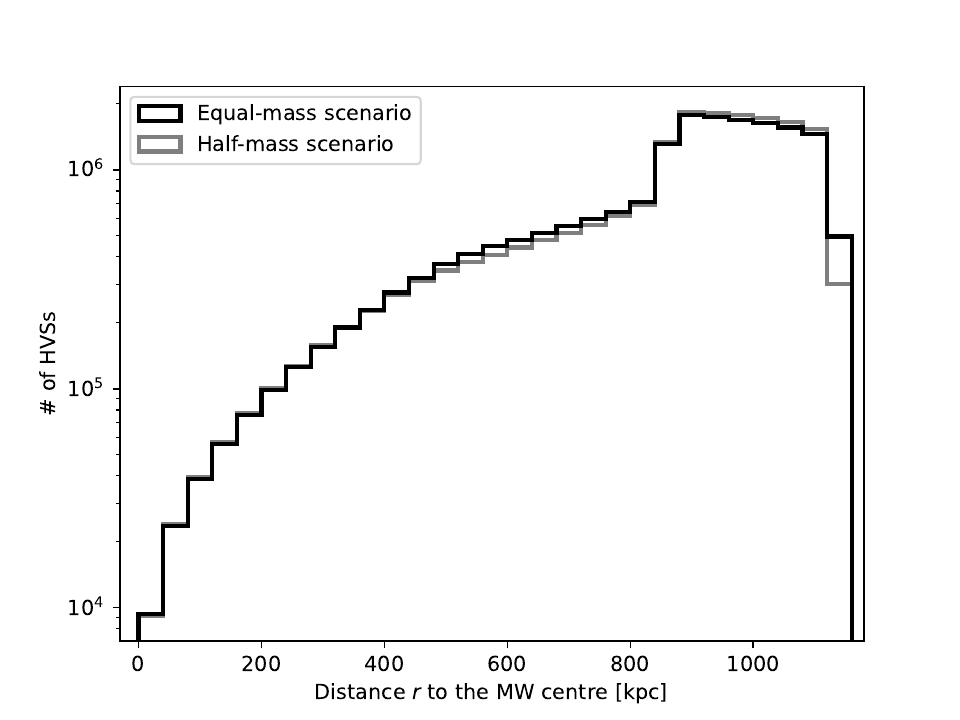}
    \caption{Distribution of minimum distance to the Milky Way centre of all simulated trajectories without any filter criteria. The star counts are displayed on a logarithmic scale. The flat peak to the right represents the HVSs that are sent off in the opposite direction to the Milky Way. Approximately $0.08$ per cent of all simulated trajectories come within $50\,\mathrm{kpc}$ of the Milky Way centre at any point during their travel time. The equal-mass scenario is shown in black, the half-mass scenario in grey. There are only small differences between the scenarios.}
    \label{fig:min_dist}
\end{figure}

The distribution of minimum distances to the MWC of all simulated HVSs in both mass scenarios is displayed in \autoref{fig:min_dist} on a logarithmic scale.
The graph shows a flat peak at the interval of distances where Andromeda is located during the $[10, 13]\,\mathrm{Gyr}$ time interval.
The peak shows the HVSs that travel in directions pointing away from the Milky Way.
For smaller distances to the Milky Way, the number of HVSs decreases rapidly.
Differences between the two mass scenarios are negligible.

\begin{figure}
	\includegraphics[width=\columnwidth]{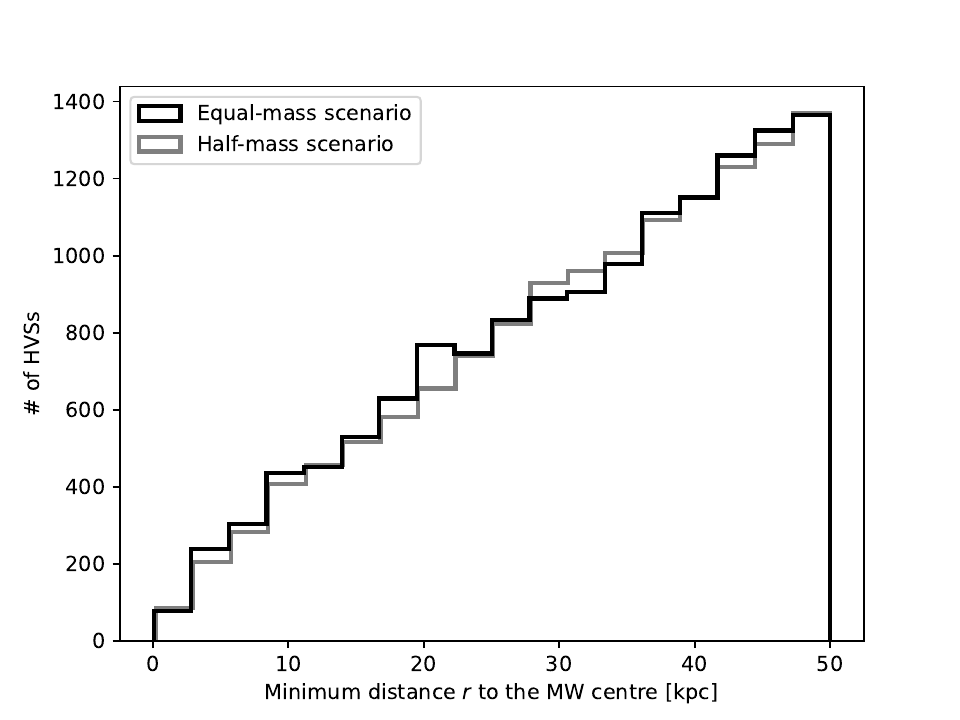}
	\includegraphics[width=\columnwidth]{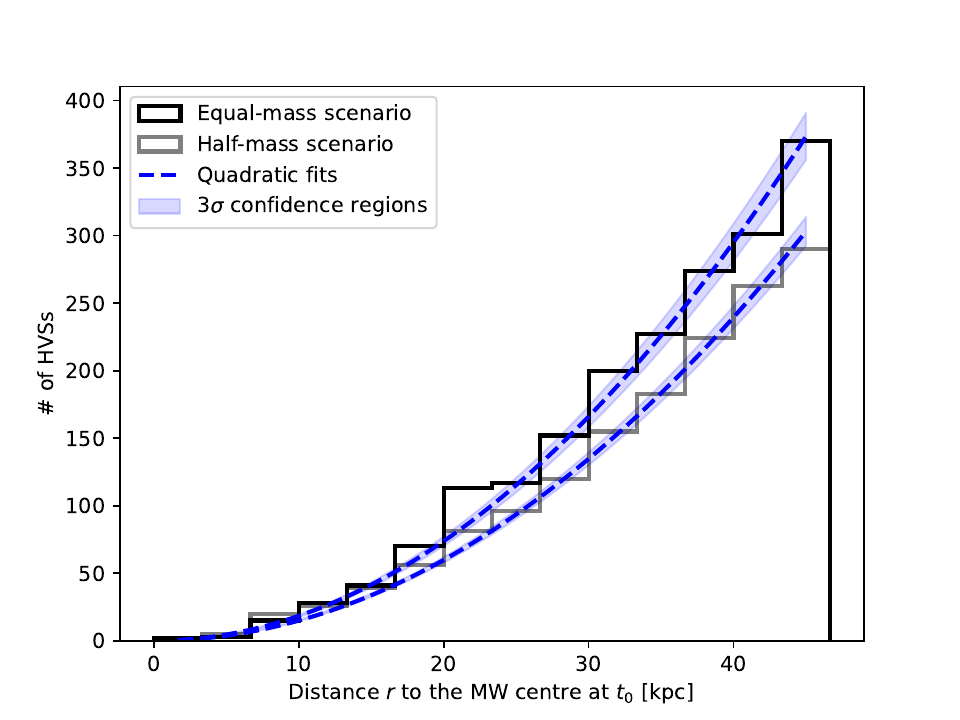}
    \caption{Distribution of distance to the Milky Way centre of simulation HVSs within $r_\text{Filter}$. The black histograms show the equal-mass scenario. The grey histograms show the half-mass scenario.  \textbf{Top:} Distances at minimum distance to the Milky Way centre. \textbf{Bottom:} Distances at present time $t_0=13.8\,\mathrm{Gyrs}$. Each scenario is fitted with a quadratic function $f(r)=a\cdot r^2$. Fits are shown as blue, dashed lines. Best-fitting values: Equal-mass ($a=0.182\pm 0.003$); Half-mass ($a=0.141\pm 0.003$). $3\sigma$ confidence regions are shown in light blue.}
    \label{fig:scenario_dist}
\end{figure}

\autoref{fig:scenario_dist} displays the distance distribution of those HVSs that reach $r_\text{Filter}$ in more detail around the MW centre and for both mass scenarios.
The top panel again shows the minimum distances while the bottom panel has the distances of HVSs within $r_\text{Filter}$ at present time.
For the overall minimum distances, both mass scenarios show the same behaviour with a linear increase of HVS counts with increasing distance to the MWC.
The linear slope indicates much lower HVS number density at higher radii.
The absolute numbers of HVSs at minimum distance for both scenarios are similar as well.

The distribution of HVSs within $r_\text{Filter}$ at present time, displayed in the bottom panel of \autoref{fig:scenario_dist}, has an overall higher HVS count for the equal-mass scenario as was explained previously.
The distributions in both scenarios show an increase in HVS counts with increasing distance to the MWC.
It is consistent with a quadratic fit $f(r)=a\cdot r^2$ in each mass scenario.
This indicates constant HVS number density within the considered $r_\text{Filter}$ sphere.
Additionally, we expect a dramatic increase of the population fraction of Andromeda HVSs with increasing distance, assuming the mass density of the Milky Way decreases with $r^{-2}$.

\begin{figure}
    \includegraphics[width=\columnwidth]{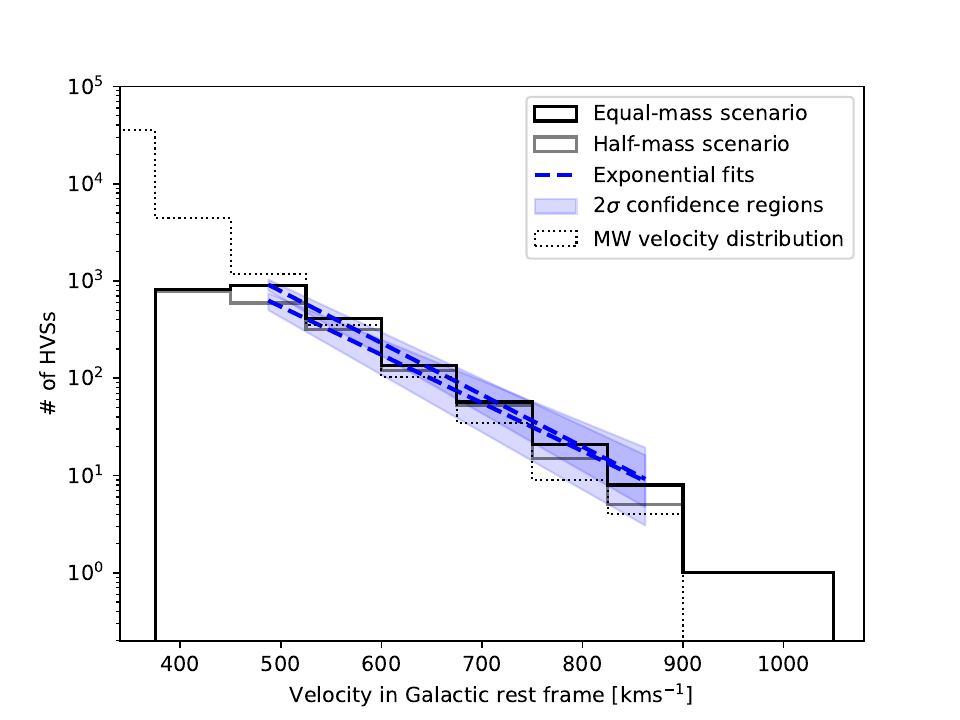}	
    \caption{Total velocity distribution of the simulated HVSs within $r_\text{Filter}$ at present time. The equal-mass scenario is shown in black, the half-mass scenario in grey. Both histograms are fitted with the exponential function from equation~(\ref{eq:fit-function}) which was also fitted to the Milky Way distribution in \autoref{fig:Marchetti}. This distribution is shown here as the dotted line. For both fits, the first and the last two bins are excluded due to incomplete information and low statistics, respectively. The fits are displayed as blue, dashed lines. $2\sigma$ confidence regions are shown in light blue.}
    \label{fig:post-Sim distr.}
\end{figure}

Next, we analyse the total velocity distributions of the simulated HVSs within $r_\text{Filter}$ at $t_0$ for both mass scenarios in \autoref{fig:post-Sim distr.}.
Both distributions are fitted with the same exponential we fitted to the Milky Way velocity distribution in \autoref{fig:Marchetti} and which we used to generate the initial HVS velocities.
The best-fitting values are reported in \autoref{tab:sim-result-fit}.
The highest velocity bins are excluded from each fit due to low statistics.
In addition, the first bin is not considered either due to incomplete information.

To start with, we find that many HVSs at lower velocities are not fast enough to be able to escape the Milky Way again.
Thus, a significant fraction of the HVSs arriving from Andromeda remain bound to the Milky Way gravity.
Overall, the HVSs still approximately follow the same exponential distribution even after the long journey to the Milky Way.
While the slopes of the result distributions seem very close to the slope of the Milky Way distribution in the logarithmic representation, a comparison with the fit values in \autoref{tab:Marchetti-fit} shows differences for each fit parameter.
However, the differences are smaller than an order of magnitude.
We note that significantly fewer data points were used for the computation of the fits of the simulation results which causes the larger uncertainties.
We also note that the simulated distributions have a similar amount of total stars in the shown bins as the initial distribution, shown as a dotted line.
This is coincidental and will change drastically based on the number of simulated HVSs in the analysis.

\begin{table}
	\centering
	\caption{Best-fitting parameters for the fitted velocity distributions of the two simulation scenarios in \autoref{fig:post-Sim distr.}.}
	\label{tab:sim-result-fit}
    Equal-mass scenario
	\begin{tabular}{cccc} 
		\hline
		Velocity scale $\sigma_v$ $\left[\mathrm{km\, s^{-1}}\right]$  & Amplitude $A$ $\left[\mathrm{km\, s^{-1}}\right]$ & $\chi^2$ & d.o.f.\\
		\hline
		\hline
		  $81.52\pm 3.05$ & $218428\pm 9295$ & $7.28$ & $4$\\
	\end{tabular}
    \vspace{0.3cm}\\
    Half-mass scenario
	\begin{tabular}{cccc} 
		\hline
		Velocity scale $\sigma_v$ $\left[\mathrm{km\, s^{-1}}\right]$  & Amplitude $A$ $\left[\mathrm{km\, s^{-1}}\right]$ & $\chi^2$ & d.o.f.\\
		\hline
		\hline
        $87.72 \pm 5.34$ & $149379 \pm 10255$ & $14.68$ & $4$\\
	\end{tabular}
\end{table}

\begin{figure}
	\includegraphics[width=\columnwidth]{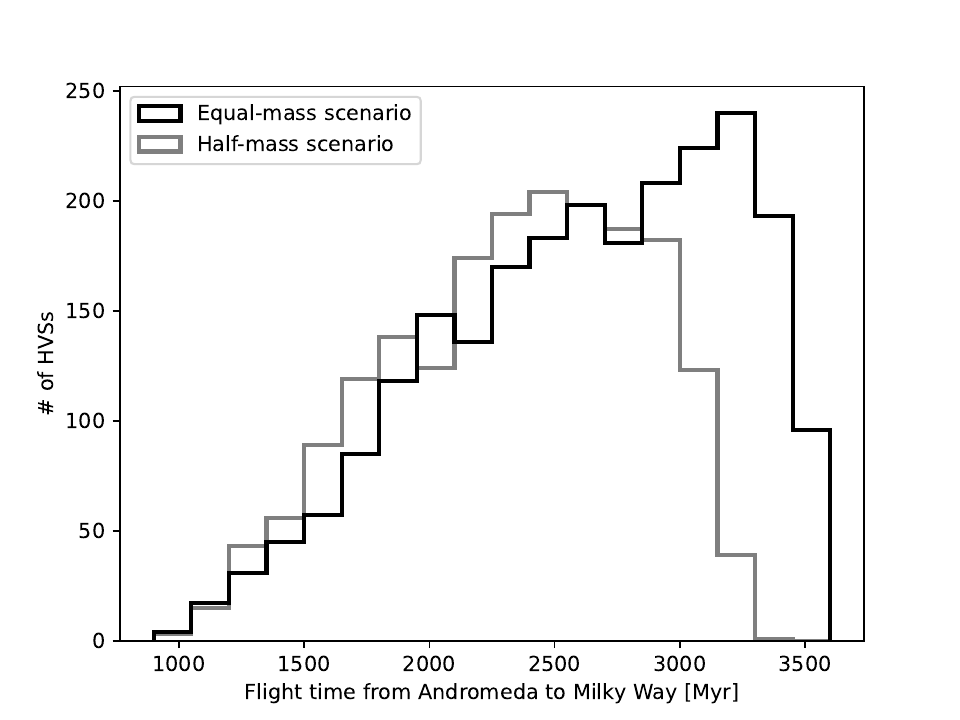}
    \caption{Distributions of flight times from Andromeda to the Milky Way centre of simulation HVSs within $r_\text{Filter}$ at present time. The black histogram shows the equal-mass scenario. The grey histogram shows the half-mass scenario. Flight times are overall higher in the equal-mass scenario.}
    \label{fig:flight-times}
\end{figure}

In \autoref{fig:flight-times}, we display the flight times of the HVSs from Andromeda to the Milky Way, from their send-off to present time.
Up to about $3000\,\mathrm{Myr}$ the distributions are similar between the two scenarios.
While the equal-mass scenario shows a large amount of HVSs with flight times up to about $3600\,\mathrm{Myr}$, the distribution in the half-mass scenario decays more quickly and only goes up to about $3300\,\mathrm{Myr}$.
The higher average initial velocity explains the lack of long flight times in the half-mass scenario.
Under the assumption that the HVSs are main sequence stars, the flight times account for a significant part of their life spans.
Particularly for the flight times of multiple Gyrs, a significant fraction of HVSs might evolve off the main sequence and not survive the journey to the Milky Way.

\begin{figure}
	\includegraphics[width=\columnwidth]{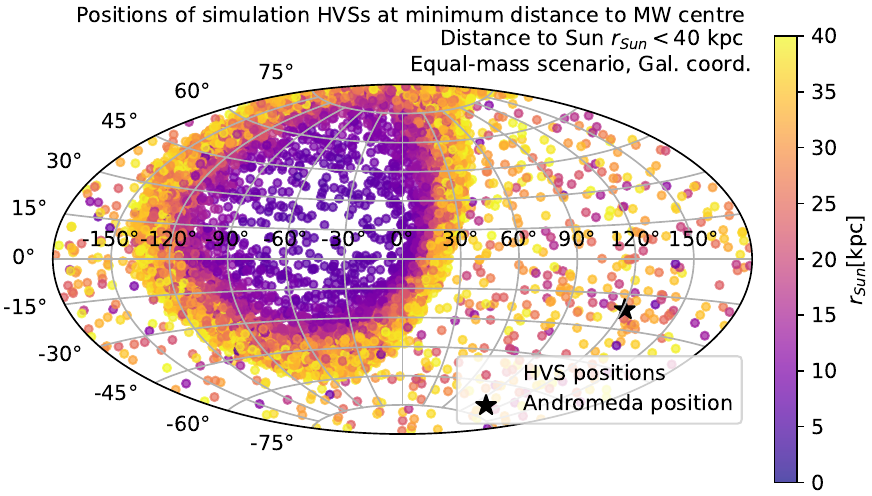}
	\includegraphics[width=\columnwidth]{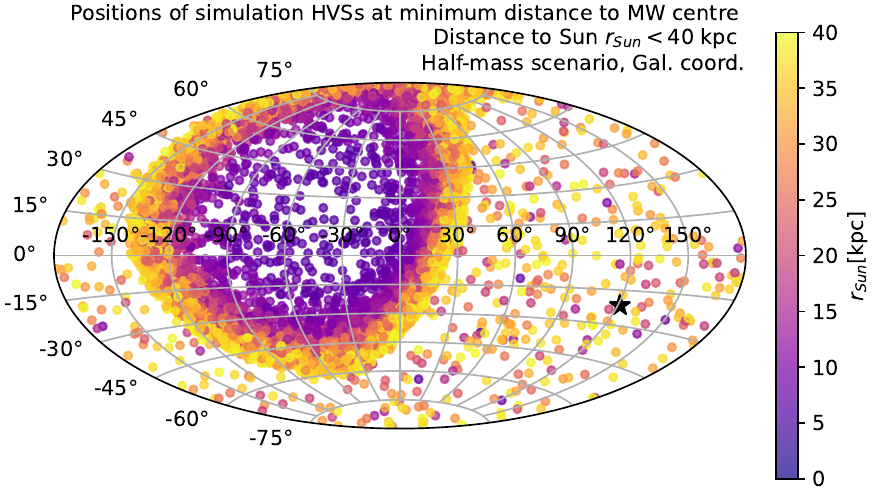}
    \caption{Skymaps showing the positions of the simulated HVSs within a sphere of radius $r=40\,\mathrm{kpc}$ around the Sun in Galactic coordinates at their minimum distance to the Milky Way centre. Colourmapped by distance to the Sun $r_\text{Sun}$. The position of Andromeda on the sky is marked by a black star. The distribution of HVSs is more concentrated in the equal-mass scenario due to the larger mass and gravitational pull of the MWC. Otherwise, the distributions are approximately identical. \textbf{Top:} Equal mass scenario. \textbf{Bottom:} Half mass scenario.}
    \label{fig:skymap-min}
\end{figure}

\begin{figure}
	\includegraphics[width=\columnwidth]{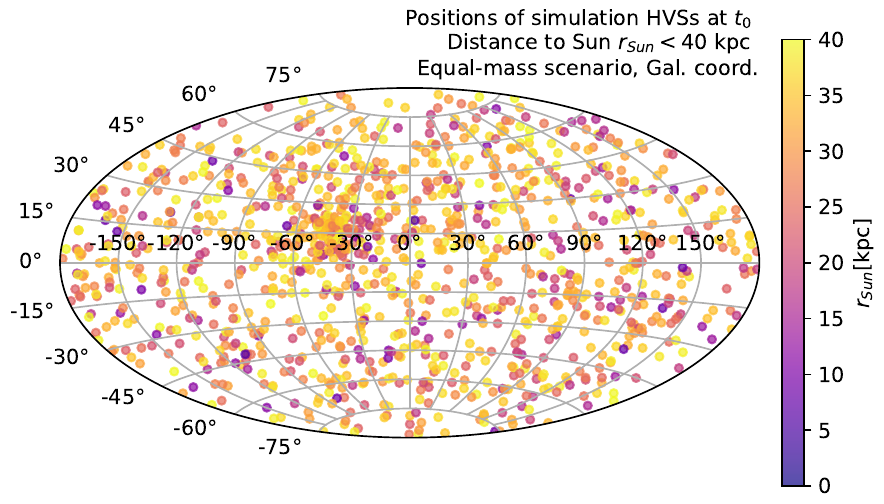}
	\includegraphics[width=\columnwidth]{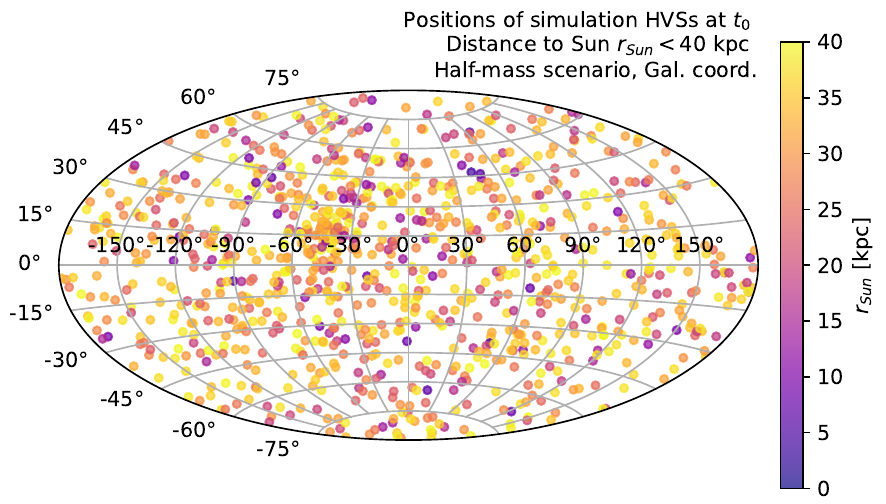}
    \caption{Skymaps showing the positions of the simulated HVSs in Galactic coordinates at present time $t_0=13.8\,\mathrm{Gyrs}$ within a sphere of radius $r=40\,\mathrm{kpc}$ around the Sun. Colourmapped by distance to the Sun $r_\text{Sun}$. The distribution is approximately homogeneous for both mass scenarios. \textbf{Top:} Equal mass scenario. \textbf{Bottom:} Half mass scenario.}
    \label{fig:skymap-present}
\end{figure}

\autoref{fig:skymap-min} and \ref{fig:skymap-present} show sky maps of the distribution of the simulated HVS positions in Galactic coordinates $(l,b)$ for both mass scenarios.
The Sun serves as the coordinate origin since this is how the HVSs would be measured from Earth.
Only HVSs within a sphere of radius $40\,\mathrm{kpc}$, that is wholly contained in the $r_\text{Filter}$ sphere, are displayed to conserve symmetry.
They are colourmapped according to their distance to the Sun.
The minimum distance distributions in \autoref{fig:skymap-min} form discs centred near the MWC at about $(-45\degr, 25\degr)$ on the opposite side of the direction of Andromeda at $(121.2\degr, -21.6\degr)$.
The discs are oriented perpendicularly to Andromeda's position vector.
Evidently, HVSs typically reach the minimum distance to the MWC just after passing it.
The HVSs that are not part the discs do not pass the MWC before present time when the simulation ends.
As such, their minimum distance positions are the same as their present time positions.

The present time positions in \autoref{fig:skymap-present} show almost isotropic distributions.
Only near the Galactic centre at about $(-30\degr, 15\degr)$, near the position of the disc centres from \autoref{fig:skymap-min}, there is a visibly higher concentration of HVSs.
This is expected due to the gravitational attraction of the MWC.
Together with the constant number density from the quadratic fits in the bottom panel of \autoref{fig:scenario_dist}, this result implies an approximately homogeneous and isotropic distribution of the simulated HVSs within the $40\,\mathrm{kpc}$ sphere around the Sun.
For both minimum distance and present time positions, the two mass scenarios do not show a significant difference.

\subsection{Comparison with \textit{Gaia} DR3 data}
\label{sec:Gaia}
To assess whether HVSs from Andromeda reaching the Milky Way is ruled out by observations, we compare the simulated HVSs at present time with star data from the \textit{Gaia} Data Release 3 (DR3) \citep{GAIA_2016, GAIA_val, GAIA_2022}.
We use the same radius of $40\,\mathrm{kpc}$ around the Sun as before and a velocity filter of $v > 500\,\mathrm{km\, s^{-1}}$ in the rest frame of the MWC.
It is important to note that only objects with measured radial velocity, full astrometric solution (sky position, parallax, proper motions) and sufficient measurement accuracy are eligible for this analysis.
Only 10721 objects of the total \textit{Gaia} catalogue fulfil these criteria.
Our methods of selecting data from the \textit{Gaia} catalogue, and the velocity coordinate transformations are detailed in Appendix~\ref{App:Gaia-data} and \ref{App:GAIA-Trafo}, respectively.

\begin{figure}
	\includegraphics[width=\columnwidth]{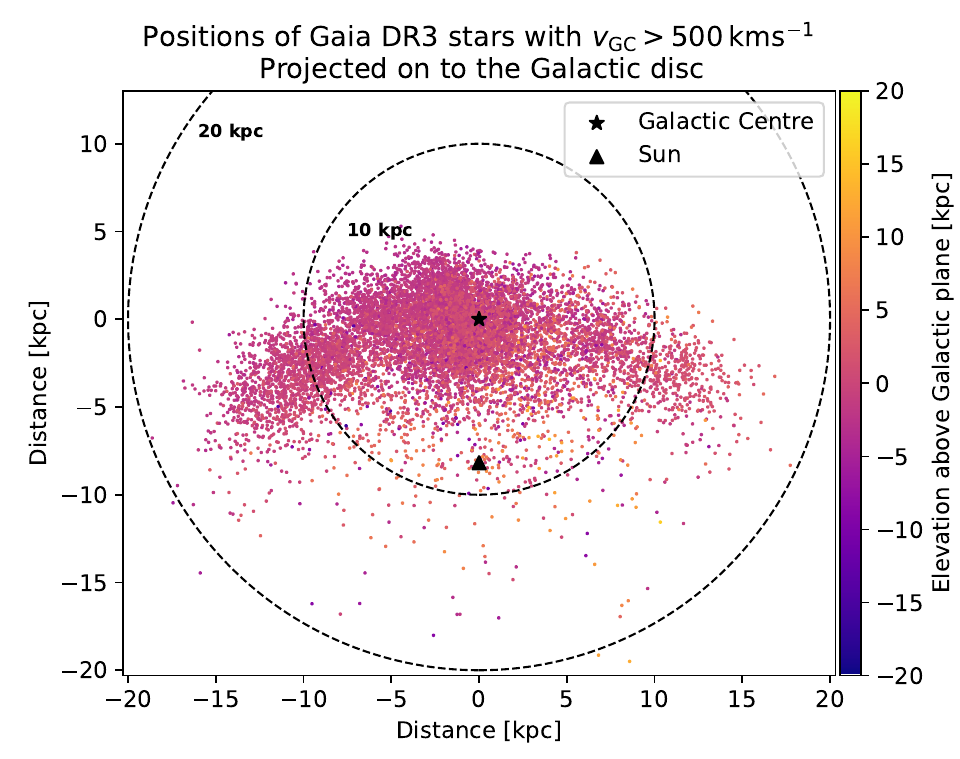}
	\includegraphics[width=\columnwidth]{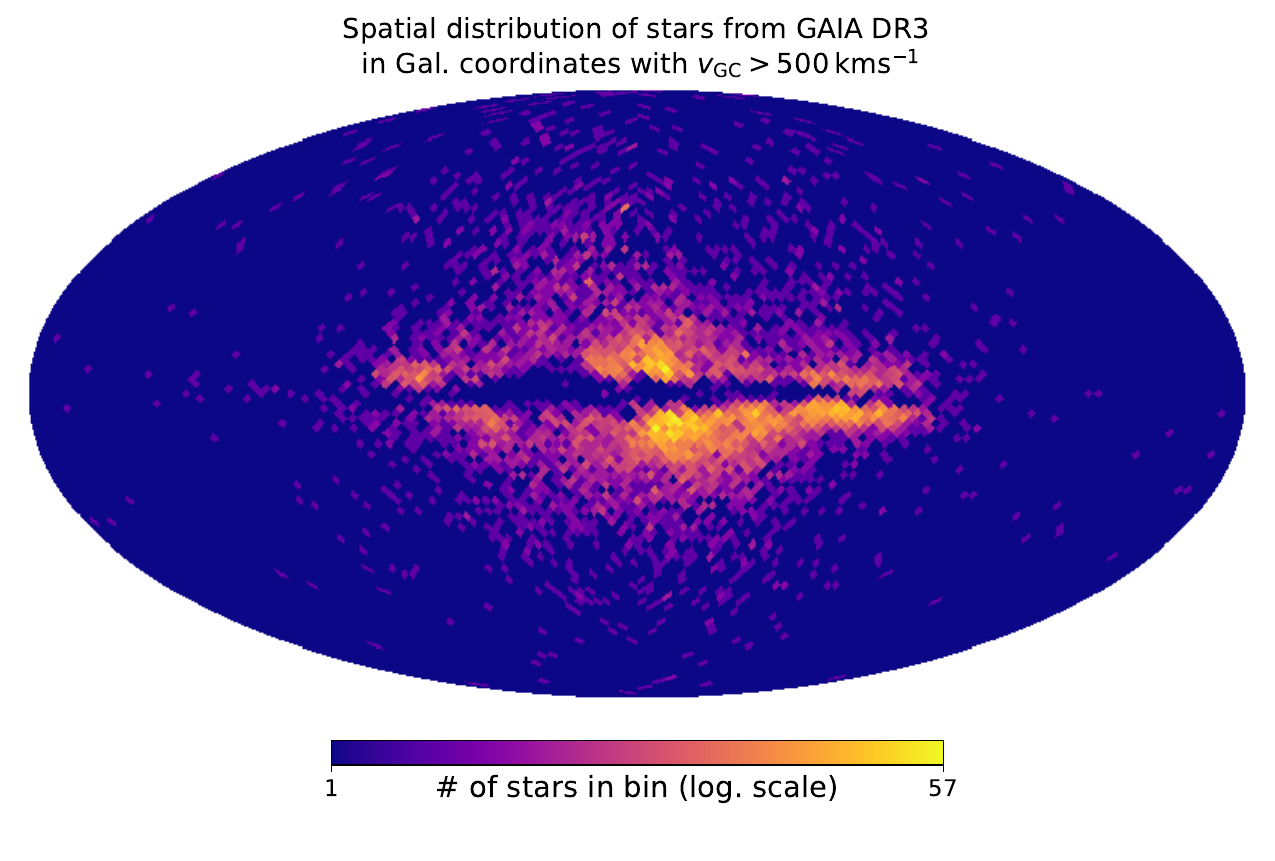}
    \caption{Positions of 10721 high-velocity in Galactic coordinates at present time $t_0=13.8\,\mathrm{Gyrs}$ within a sphere of radius $r=40\,\mathrm{kpc}$ around the Sun. Data are taken from Data Release 3 of the \textit{Gaia} satellite mission. \textbf{Top:} Star positions projected on to the Galactic plane and colourmapped by their elevation above the disc. \textbf{Bottom:} Spatial distribution of \textit{Gaia} stars as a population density sky map.}
    \label{fig:GAIA skymap}
\end{figure}

In \autoref{fig:GAIA skymap}, we compare the positional data.
The top panel shows the star positions, projected on to the Galactic plane with their elevation above it shown as colour. 
Almost all shown stars are close to zero elevation.
The majority of stars is contained within an arc shape that is densest around the MWC.
The area outside the arc is only sparsely populated.
We expect a large concentration of high-velocity stars around the MWC and the Galactic disc due to the high stellar density in this region.
In the bottom panel, the population density on the sky is displayed.
We notice a prominent feature directly on the Galactic disc where the number of stars seems unexpectedly low.
This is due to inaccurate measurements caused by dust attenuation along the line of sight, and the corresponding data cuts. 
Comparing this panel with \autoref{fig:skymap-present}, we see that \textit{Gaia} observes stars at the region opposite Andromeda where we expect a concentration of HVSs from the simulation results.
Overall, the distribution is very different from the isotropic scatter of the simulated HVSs.

\begin{figure}
	\includegraphics[width=\columnwidth]{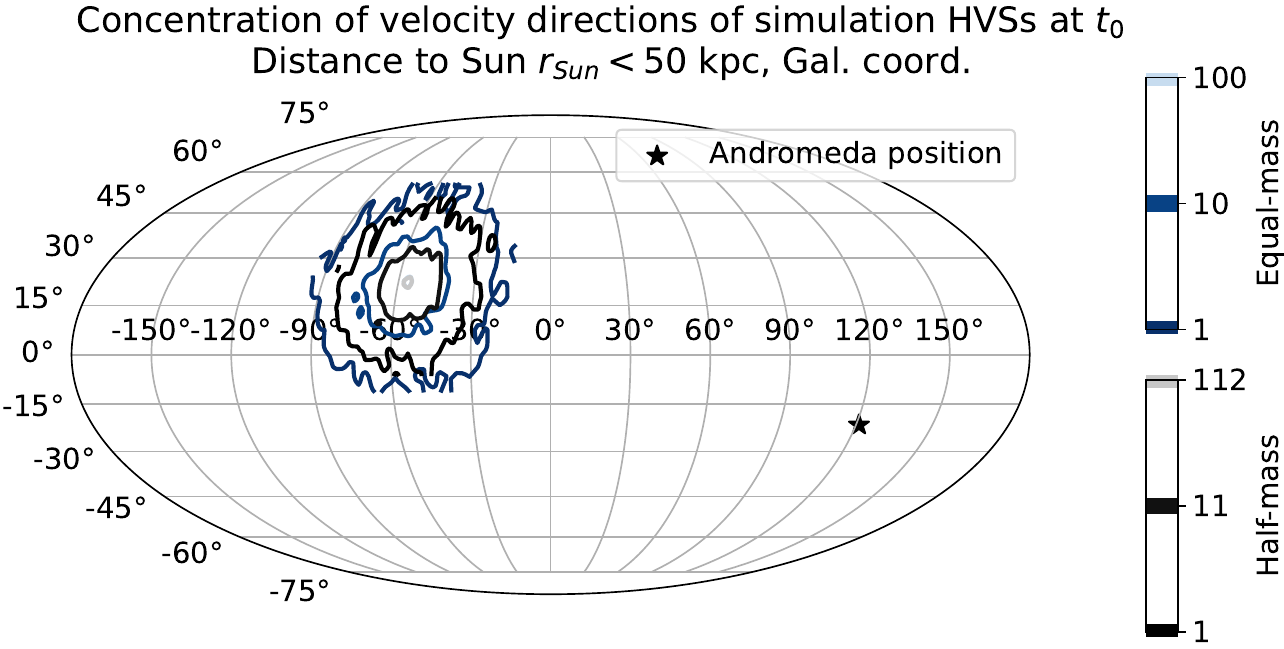}
	\includegraphics[width=\columnwidth]{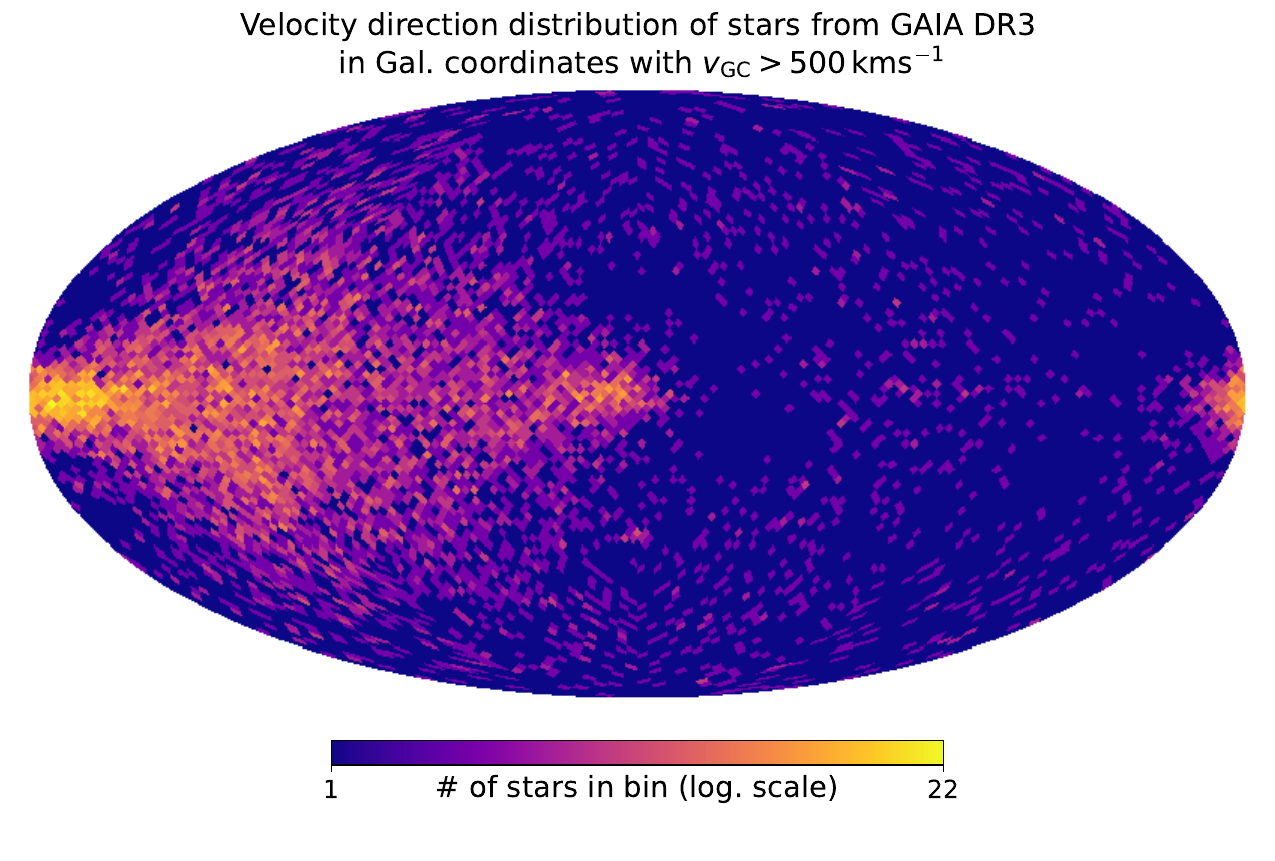}
    \caption{Comparison of velocity directions from simulation results and \textit{Gaia} DR3 data. Directions are shown as if the velocity vector passed through the coordinate origin. Full $50\,\mathrm{kpc}$ radius around the MWC is considered. \textbf{Top:} Distribution of velocity directions of simulated HVSs in the MWC rest frame as contour plots. Blue contours show the broader distribution in the equal-mass scenario, grey-scale contours the distribution in the half-mass scenario. The velocity directions are more densely concentrated in the half-mass scenario. \textbf{Bottom:} Velocity directions of \textit{Gaia} stars with velocity in the MWC rest frame $v_\text{MWC}>500\,\mathrm{km\, s^{-1}}$ as a population density sky map.}
    \label{fig:GAIA vel skymap}
\end{figure}

Further, we directly compare the HVS velocity directions from the simulation results in the MWC rest frame with the corresponding \textit{Gaia} data in \autoref{fig:GAIA vel skymap}.
The simulation results in the top panel show a narrow range of possible directions which, expectedly, lie opposite of the position of Andromeda on the sky.
The distribution of HVS directions in the equal-mass scenario is broader than in the half-mass scenario while also being less concentrated in the center.
This is caused by stronger gravitational focusing of the trajectories in the equal-mass scenario, meaning the trajectories that reach the Milky Way in the half-mass scenario, on average, point more directly from Andromeda at the Milky Way.

In comparison, the \textit{Gaia} HVS velocity vectors in the bottom panel, shown as a logarithmic population density map, have strong preference along the position vector of the MWC $(0\degr, 0\degr)$ and the opposite direction $(180\degr, 0\degr)$.
Further, we note a less populated preference in the directions of negative Galactic longitude, and Galactic latitude angles between $60\degr$ and $-60\degr$.
This preference is denser within $([-150, -90]\degr, [-30, 30]\degr)$.
The centre of this area lies opposite to the movement direction of the Sun.
It also overlaps with the directions of the simulated HVSs.
While we cannot observe clustering around the direction pointing away from Andromeda as in the simulations, we cannot exclude that some of the observed HVSs are indeed coming from Andromeda.

From these two comparisons, we conclude that the Andromeda HVS scenario is not excluded by \textit{Gaia} data.
In addition, the strongly focused velocity directions of HVSs from Andromeda might allow for their detection if the HVSs numbers are large enough to cause a feature in a plot like the bottom panel of \autoref{fig:GAIA vel skymap}.

\subsection{How many Andromeda HVSs are in the Milky Way?}
\label{sec:HVS-amount}
Using the constraints of the ejection rate of HVSs from the MWC from \citet{Evans_20222}, we can estimate the number of HVSs from Andromeda we expect in the Milky Way at present time.
However, we need to assume the equal-mass scenario and that the ejection rate stays approximately the same between $10$--$13\,\mathrm{Gyr}$ after the Big Bang.
We simply multiply the ejection rate per year with this time interval and the fraction of HVSs we expect from Andromeda (see \autoref{tab:resultdist}).
The results of the estimation are shown in \autoref{tab:ejection-rate}.
We acknowledge that these numbers might be lower since not all HVSs survive the travel time to the Milky Way.
The significant differences between the two estimates are due to model dependencies that are discussed in \citet{Evans_20222}.
Even for the lower bound of the ejection rate, we expect an amount of HVSs that is greater than ten.
Such an amount is unlikely to be detected since \textit{Gaia} DR3 only has full position and velocity information for a small fraction of its catalogue.
The large amount of HVSs in the upper bound has a much higher likelihood to appear in measurements.
However, a more detailed search for Andromeda HVSs in measurements of Milky Way stars is beyond the scope of this work.
\begin{table}
	\centering
	\caption{Estimations of the amount of Andromeda HVSs expected to be in the Milky Way at present time based on constraints on the ejection rate of HVSs from the MWC from \citet{Evans_20222}. Only the equal-mass scenario is shown. The top line displays the lower bound, the bottom line the upper bound.}
	\label{tab:ejection-rate}
	\begin{tabular}{ccc} 
		\hline
		HVS ejection rate [$\mathrm{yr}^{-1}$] & Andromeda HVSs expected in the Milky Way\\
		\hline
		\hline
		$10^{-4.5}$ & $12$ \\
		$10^{-2}$ & $3910$ \\
		\hline
		&  & \\
		&  & \\
	\end{tabular}
\end{table}

For the Hills mechanism, \citet{Sherwin_2008} expect there to be of order 1000 HVSs within the virialised Milky Way halo.
We consider that this work does not take the entire Milky Way halo into account, but only about $1.43\%$ of the halo of radius $R_\text{M,200}=206\,\mathrm{kpc}$ assumed in \citet{Sherwin_2008}.
Compared to the result obtained in this work, the lower bound and values up to about 150 HVSs are compatible.
Notable differences between the two approaches include the models for the mass distribution and dynamics of the Local Group.
Further, \citet{Sherwin_2008} consider the stellar masses of the ejected HVSs, which we neglect.

As we briefly discussed in Section \ref{sec:maths}, we also considered the power law HVS velocity distribution from \citet{Sherwin_2008} \textbf{for} our model. 
For this comparison, we again simulated $1.8 \cdot 10^8$ trajectories in the equal-mass scenario, now with initial velocities generated from this power-law velocity distribution.
We find that we expect about $0.0078$ per cent of HVSs to reach the $50\,\mathrm{kpc}$ radius around the MWC.
The resulting lower and upper bounds for the total amount of HVSs are presented in \autoref{tab:Sherwin-comp}.
The amount of HVSs expected in the Milky Way using the power law distribution remains within the same order of magnitude for both the upper and lower bound.
As such, the conclusion does not change.
\begin{table}
	\centering
	\caption{Number of Andromeda-ejected HVSs presently in Milky Way according to the power law initial velocity distribution of \citet{Sherwin_2008}. The top line displays the lower bound, the bottom line the upper bound.}
	\label{tab:Sherwin-comp}
	\begin{tabular}{ccc} 
		\hline
		HVS ejection rate [$\mathrm{yr}^{-1}$] & HVS amount predicted by power law\\
		\hline
		\hline
		$10^{-4.5}$ & $7$ \\
		$10^{-2}$ & $2341$ \\
		\hline
		&  & \\
		&  & \\
	\end{tabular}
\end{table}
\section{Conclusions}
\label{sec:Concl}
To find out whether hypervelocity stars (HVS) from Andromeda can reach the Milky Way, we have developed a simulation model of the gravitational system of the two galaxies.
Using this model, we have calculated the trajectories of $1.8\cdot 10^7$ hypervelocity stars each for two different scenarios.
We have considered two mass scenarios with the Milky Way and Andromeda having equal mass, and the Milky Way having about half of Andromeda's mass, respectively.

The HVS initial positions have been randomly generated near the centre of Andromeda and ejected in random directions with initial velocity magnitudes based on Milky Way star behaviour and Andromeda's escape velocity.
We have found that $0.013$ and $0.011$ per cent of simulated HVSs are within a radius of $50\,\mathrm{kpc}$ around the Milky Way centre (MWC) at present time.

We have analysed the distance distributions within this $50\,\mathrm{kpc}$ filter radius.
The minimum distances to the MWC follow a linear distribution.
This means the number density decreases significantly at large radii.
The mass scenarios show no significant differences in this regard.
In contrast, for the present time results, the equal-mass scenario shows a higher count of HVSs due to the stronger gravitation of the Milky Way and the initial HVS velocities in this scenario.
The present time distance distributions show parabolic behaviour and have been fitted with a quadratic function, indicating homogeneous HVS number density within the considered sphere.

We have compared the velocity distributions of the present time data sets in both mass scenarios to the data set we used for the generation of initial velocities.
We have found that, even after the long journey to the Milky Way, the HVSs still approximately follow the initial velocity distribution.
Some HVSs from Andromeda slow down so much that they end up bound to the Milky Way.
Additionally, we have considered the HVS flight times and found that a significant fraction of the ejected HVSs might evolve off the main sequence during the journey to the Milky Way.

From the simulation results, we have displayed as sky maps all HVSs arriving within $40\,\mathrm{kpc}$ of the Sun at present time $t_0$ as well as all HVSs that reach this minimum distance at any point during their trajectory.
The positions of HVSs at their minimum distance form a disc centred on the MWC, opposite of Andromeda.
The present time positions are distributed nearly isotropically and homogeneously within the considered sphere around the Sun.
They are, expectedly, slightly more concentrated around the MWC due to its gravitational attraction.
The mass scenarios show negligible differences for both minimum distance and present time sky maps.

We have compared this result with high-velocity star positions from the \textit{Gaia} Data Release 3.
We have found that the simulated HVS star positions are consistent with the spatial distribution of observed HVSs, taking into account that HVSs should also originate from the MW itself, an effect that is not included in our simulations.
We have compared the velocity directions from the simulation results with the \textit{Gaia} data to find an approximately homogeneous population of velocity directions that overlaps with the simulated velocity directions.

In addition, we have estimated the expected amount of Andromeda HVSs in the Milky Way at present time using constraints on the ejection rate of HVSs from the MWC.
We have found that even for the lower bound, we expect an amount greater than ten.
The upper bound results in a large number of Andromeda HVSs that might be detectable in a more detailed analysis.
We have compared these results to earlier work and found that they are compatible. In addition, we have tested our model with initial velocities generated from the HVS velocity distribution from said earlier work. As a result, the expected number of Andromeda HVSs only changes within an order of magnitude.

We conclude that it is possible for HVSs from Andromeda to travel towards and reach the Milky Way.
They are expected to form an approximately isotropic and homogeneous distribution around the MWC.
Their expected velocity directions point away from Andromeda. Only a small fraction of HVSs ejected from Andromeda are expected to migrate to the Milky Way.
However, it might be possible to detect them based on their velocity and trajectory orientation.
Further, it would be interesting to extend this analysis to include estimates of the stellar ages and other astrophysical information of HVSs in order to gain further insight into their origin and migration history.

\section*{Acknowledgements}
We acknowledge valuable discussions with Marius Neumann, Ferdinand J{\"u}nemann, David Clarke, Jelena K{\"o}hler and Pranav Sampathkumar during the preparation of this paper.

The computations in this work were performed on the GPU cluster at Bielefeld University.
We thank the Bielefeld \textit{HPC.NRW} team for their support.

MF was supported by the United Kingdom STFC Grant ST/X000753/1

This work was co-funded by the Erasmus+ programme of the European Union.

This work has made use of data from the European Space Agency (ESA) mission {\it Gaia} (\url{https://www.cosmos.esa.int/gaia}), processed by the {\it Gaia} Data Processing and Analysis Consortium (DPAC, \url{https://www.cosmos.esa.int/web/gaia/dpac/consortium}).
Funding for the DPAC has been provided by national institutions, in particular the institutions participating in the {\it Gaia} Multilateral Agreement.

This work made use of \textsc{astropy} (\url{http://www.astropy.org}), a community-developed core Python package for Astronomy \citep{astropy:2013, astropy:2018}, \textsc{numpy} \citep{Harris_2020}, \textsc{scipy} \citep{2020SciPy-NMeth}.
All figures in this work were produced using \textsc{matplotlib} \citep{Hunter_2007}, \textsc{healpix} \citep{Gorski_2005} and \textsc{topcat} \citep{2005Topcat}.

\section*{Data Availability}
The simulation code and the simulation results are available as supplementary material and can also be found on github (\url{https://github.com/lguelzow/On-Stellar-Migration-from-the-Andromeda-Galaxy}).

The data from \citet{Marchetti_2021} for the velocity analysis of Milky Way stars is available on T. Marchetti's website (\url{https://sites.google.com/view/tmarchetti/research}).

The acquisition of the \textit{Gaia} Data Release 3 data is described in Appendix~\ref{App:Gaia-data}.



\bibliographystyle{mnras}
\bibliography{bibliography} 





\appendix

\section{Data Selection from \textit{Gaia} DR3 Catalogue}
\label{App:Gaia-data}
The \textit{Gaia} Data Release 3 data are provided by the \textit{Gaia} archive website at \texttt{https://gea.esac.esa.int/archive/}.
We use the following prompt in the ADQL interface:

\begin{verbatim}
SELECT source_id, parallax, ra, dec, pmra, pmdec,
radial_velocity, parallax_error, ra_error, dec_error,
pmra_error, pmdec_error, radial_velocity_error,
phot_g_mean_mag, nu_eff_used_in_astrometry, 
pseudocolour, ecl_lat, astrometric_params_solved
FROM gaiadr3.gaia_source
WHERE parallax >= 0.02
AND ruwe < 1.4
AND rv_nb_transits > 10
AND rv_expected_sig_to_noise >= 5
AND abs(parallax) > 2 * parallax_error
AND abs(pmra) > 5 * pmra_error
AND abs(pmdec) > 5 * pmdec_error
AND abs(radial_velocity) > 5 * radial_velocity_error
AND sqrt(power(radial_velocity, 2)
+ power((pmra / parallax) * 4.744213026, 2)
+ power((pmdec / parallax) * 4.744213026, 2)) >= 150
\end{verbatim}

The \texttt{SELECT} line provides the columns of the data base used for the result table.
The \texttt{FROM} line sets the data base from which the data are drawn.
The \texttt{WHERE} and \texttt{AND} lines refer to further conditional statements for the data selection.
Here, we use multiple conditions to ensure we obtain pure and accurate data from the \textit{Gaia} catalogue.
First, we require relative errors smaller than $20\%$ for radial velocity and proper motion (\texttt{pmra} and \texttt{pmdec}).
For parallax measurements, we initially use a more relaxed criterion of relative errors smaller than $50\%$.
After this selection we remove the parallax offset present in the \textit{Gaia} catalogue as detailed in \citet{Lindegren2020} and \citet{Groeneweg2021}.
The parameters \verb|phot_g_mean_mag|, \verb|nu_eff_used_in_astrometry|, \verb|pseudocolor|, \verb|ecl_lat| and \verb|astrometric_params_solved| are required for this.
Once the parallax values are corrected, we apply the criterion of relative error smaller than $20\%$.
This condition matches the parallax accuracy to the rest of the data and enables us to accurately estimate distances from parallaxes by simply inverting them \citep{Bailer-Jones2015}.
In addition to accuracy conditions on the parameters we use directly, we also select stars with criteria on the parameters \verb|ruwe| (renormalised unit weight error), \verb|rv_nb_transits| (number of epochs used to determine the radial velocity of each star) and \verb|rv_expected_sig_to_noise| (expected signal-to-noise ratio in the combination of individual spectra used to determine the \textit{Gaia} DR3 radial velocity).
For details on these criteria, we refer to \citet{Marchetti_2022} from where we have adopted them.
Lastly, the final condition calculates the velocity magnitude of the object from its radial velocity and proper motion and only adds it to the result table if it is $\geq 150\,\mathrm{km\, s^{-1}}$.
Notably, this velocity magnitude is in the heliocentric rest frame.
We choose this low threshold to make sure we do not miss any stars in the catalogue that have a velocity larger than $500\,\mathrm{km\,s^{-1}}$ in the Galactocentric rest frame.

\section{Coordinate Transformation of the \textit{Gaia} DR3 Velocities}\label{App:GAIA-Trafo}

The data from the \textit{Gaia} Data Release 3 catalogue are given in the equatorial coordinate system.
It utilises the celestial sphere of the Earth.
We use the \textsc{AstroPy} function \textsc{Skycoord} to transform the star positions from the catalogue to Galactic coordinates.
In order to obtain the distance to the Sun $d_S$  and velocity vectors $\mathbfit{v}_\text{Gaia}$ of the measured HVSs, we use the given parallax $P$ and radial velocity $v_\text{rad}$ as well as the proper motion velocities $\omega_\text{RA}, \omega_\text{Dec}$ in the direction of the equatorial coordinate components, right ascension ($\text{RA}$) and declination ($\text{Dec}$).
The parallax $P$ in milliarcseconds (mas) is simply converted to the distance to the Sun $d_S$ in kiloparsec (kpc) by taking the inverse.

\begin{figure}
\centering
\includegraphics[width=\columnwidth]{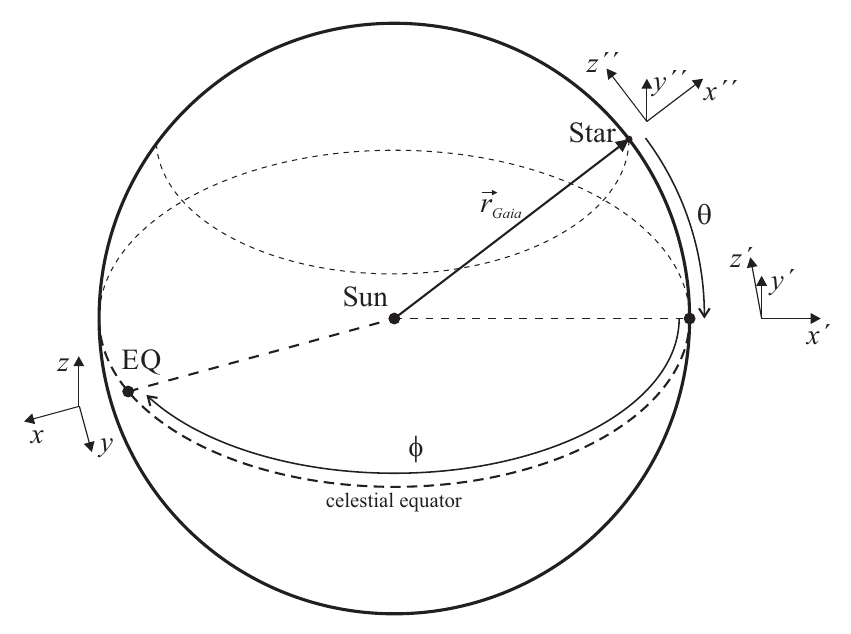}
\caption{Schematic illustration of the transformation of the \textit{Gaia} catalogue velocity data in hour angle equatorial coordinates.
We transform the coordinate system $(x^{\prime\prime},y^{\prime\prime},z^{\prime\prime})$ determined by the position vector $\mathbfit{r}_\text{Gaia}$ of the HVS by rotating by $\theta=\text{Dec}$ around the $y^{\prime\prime}$-axis and by $\phi=-\text{RA}$ around the $z^\prime$-axis of the transitional coordinate system $(x^\prime,y^\prime,z^\prime)$.
The $z^\prime$-axis is slightly tilted for visual clarity.
Now it matches the coordinate system $(x,y,z)$ at the position $(0\degr, 0\degr)$, the vernal equinox (EQ).
This coordinate system is aligned with the standard equatorial coordinate frame and gives us the velocity vector $\mathbfit{v}_\text{Gaia}$.}
\label{GAIA-trafo}
\end{figure}

To find the velocity vector of a given object in Galactic coordinates, we have to perform multiple coordinate transformations since we only have the previously mentioned measurement values.
First, we determine the velocity vector of a HVS in equatorial Cartesian coordinates from the \textit{Gaia} data using two coordinate rotations.
This process is schematically displayed in \autoref{GAIA-trafo}.
After these rotations and a transformation to spherical coordinates, we can again use the \textsc{Skycoord} function to convert the calculated velocity direction from equatorial coordinates to Galactic coordinates. 

Initially, we construct the velocity vector $\mathbfit{v}\,^{\prime\prime}_\text{Gaia}$ of a \textit{Gaia} object in a Cartesian coordinate frame $(x^{\prime\prime},y^{\prime\prime},z^{\prime\prime})$ that uses the position vector $\mathbfit{r}_\text{Gaia}$ of the HVS in equatorial Cartesian coordinates as its $x^{\prime\prime}$-axis and the equatorial coordinate directions $\text{RA}$ and $\text{Dec}$ at the coordinates of the HVS as its $y^{\prime\prime}$- and $z^{\prime\prime}$-axes, respectively (see \autoref{GAIA-trafo}, top right).
The velocity component $v^{\prime\prime}_{\text{Gaia},x}$ along the $x^{\prime\prime}$-axis is simply the radial velocity of the HVS.
We calculate the $y^{\prime\prime}$- and $z^{\prime\prime}$-components using the parallax and proper motion of the HVS with

\begin{equation}
\begin{aligned}
v^{\prime\prime}_{\text{Gaia},y}&=\frac{\omega_\text{RA}}{P}\cdot\text{cos}(\text{Dec})\cdot C\\
v^{\prime\prime}_{\text{Gaia},z}&=\frac{\omega_\text{Dec}}{P}\cdot C.
\end{aligned} 
\end{equation}

We need to multiply by $\text{cos}(\text{Dec})$ in $v^{\prime\prime}_{\text{Gaia},y}$ since proper motion in the horizontal direction near the North pole of a spherical coordinate frame covers more real distance than near the equator of the coordinate frame in the same time frame.
This is already included in the \textit{Gaia} catalogue value of $\omega_\text{RA}$.
The constant $C\approx 4.744$ is needed for the conversion to $\mathrm{km\, s^{-1}}$.

To find the velocity vector $\mathbfit{v}_\text{Gaia}$ in the equatorial Cartesian frame $(x,y,z)$ aligned with the standard right-handed equatorial coordinate frame, we rotate the coordinate frame $(x^{\prime\prime},y^{\prime\prime},z^{\prime\prime})$ to match $(x,y,z)$.
This action is equivalent to rotating $\mathbfit{v}\,^{\prime\prime}_\text{Gaia}$ with transposed rotation matrices.
We use the transpose of the $y$-axis rotation matrix $\textbf{R}^T_y$  to rotate around the $y^{\prime\prime}$-axis and the corresponding matrix $\textbf{R}^T_z$ for the rotation around the $z^\prime$-axis of the resulting intermediate coordinate frame $(x^{\prime},y^{\prime},z^{\prime})$ (see \autoref{GAIA-trafo}, centre right)
\begin{equation}
\begin{aligned}
\textbf{R}^T_y &=
\begin{pmatrix}
\text{cos}\,\alpha & 0 & -\text{sin}\,\alpha\\
0 & 1 & 0\\
\text{sin}\,\alpha & 0 & \text{cos}\,\alpha
\end{pmatrix}\quad\text{and}\\
\textbf{R}^T_z &=
\begin{pmatrix}
\text{cos}\,\beta & \text{sin}\,\beta & 0\\
-\text{sin}\,\beta & \text{cos}\,\beta & 0\\
0 & 0 & 1
\end{pmatrix}.
\end{aligned}
\end{equation}
In this definition, $\alpha$ and $\beta$ are arbitrary angles.
We use the corresponding position coordinates $\mathbfit{r}_\text{Gaia}=(\text{RA}, \text{Dec})$ as rotation angles.
We rotate first around the $y^{\prime\prime}$-axis by the angle $\alpha=\text{Dec}$ and subsequently around the $z^{\prime}$-axis by the angle $\beta=-\text{RA}$.
We need to to use the negative of $RA$ because the data are given in the hour angle system which is the left-handed equivalent to the standard equatorial frame.

Now, we can simply perform the transformation from Cartesian to spherical coordinates to find the point on the sphere in equatorial coordinates the velocity vector points at if placed in the coordinate origin.
We transform the velocity direction into the Galactic coordinate frame using \textsc{Skycoord}.

\bsp	
\label{lastpage}
\end{document}